\documentstyle[epsfig]{aipproc}

\begin{document}

\title{The Lyman Alpha Forest in Hierarchical Cosmologies} 
 
\author{M. Machacek$^1$, G.L. Bryan$^2$, P. Anninos$^3$,
A. Meiksin$^4$, M.L. Norman$^5$, and Y. Zhang$^3$}
\address{$^1$Physics Department, Northeastern University, Boston, MA 02115\\
$^2$Physics Department, Massachusetts Institute of Technology, Cambridge, MA
02139\\$^3$Laboratory of Computational Astrophysics, National Center for
Supercomputing Applications, 405 Matthews Ave, Urbana, IL 61801\\
$^4$Institute of Astronomy, University of Edinburgh, Royal Observatory,
Blackford Hill, Edinburgh EH9 3HJ, UK\\$^5$Astronomy Department, University of
Illinois at Urbana-Champaign, Urbana, IL 61801}

\maketitle

\begin{abstract}
The comparison of quasar absorption spectra with numerically simulated spectra
from hierarchical cosmological models of structure formation promises to be a 
valuable tool to discriminate among these models. We present simulation 
results for the column density, Doppler $b$ parameter, and optical depth 
probability distributions for five popular cosmological models.
\end{abstract}

\section*{Introduction}

A physical picture of the Ly$\alpha$ forest in hierarchical cosmologies has
recently emerged from numerical simulations \cite{sims,bryan} in which the 
absorbers that give rise to low column density lines 
($N_{HI} < 10^{15}$~cm$^{-2}$ at $z \sim 3$) are large, unvirialized objects
with sizes of $\sim 100$~kpc and densities comparable to the cosmic mean. 
Since the absorbers grow from the primordial density fluctuations through 
gravitational amplification, statistics of the forest may be used to test 
various models of structure formation.  We discuss the numerical stability 
of the 
statistics against changes in simulation box size and spatial resolution in 
detail elsewhere. \cite{bryan}  We focus here on examples 
from our model comparison study \cite{machacek} in which statistics of the  
Ly$\alpha$ forest are computed in five cosmological models: 
the standard cold dark matter model (SCDM), a flat cold dark matter model 
with nonvanishing cosmological constant (LCDM), a low density cold dark 
matter model (OCDM), a flat cold dark matter model with a tilted power 
spectrum (TCDM), and a critical model with both cold dark matter and two 
massive neutrinos (CHDM). The initial fluctuations, assumed to be Gaussian, 
are normalized using $\sigma_{8h^{-1}}$ to agree with the observed 
distribution of clusters of galaxies, although all but SCDM are also 
consistent with the COBE measurements of the cosmic microwave background. 
By varying $\sigma_{8h^{-1}}$ within 
a given model (SCDM), we also investigate the dependence of the statistics on 
changes in the fluctuation power spectrum.

The simulation technique uses a particle-mesh algorithm to follow the dark
matter and the piecewise parabolic method~\cite{bryan95} to simulate gas 
dynamics.  The simulation box length is $9.6$~Mpc (comoving) with spatial 
resolution of $18.75h^{-1}$ ($37.5h^{-1}$)~kpc for the model (power) 
comparison studies, respectively. Nonequilibrium effects are followed 
\cite{anninos} for 
six particle species (HI, HII, HeI, HeII, HeIII, and the electron density). 
We assume a spatially-constant radiation field computed from the observed QSO
distribution \cite{haardt} which reionizes the universe around $z \sim 6$ 
and peaks at $z \sim 2$.  Synthetic spectra are generated along $300$ random 
lines of sight through the volume, including the effects of peculiar velocity 
and thermal broadening.  The spectra are normalized to give a mean optical 
depth $< \tau > = 0.3$ at $z = 3$ to agree with observation. We do not include
the effects of radiative transfer, self-shielding or star formation and so 
can not address the physics of the highest column density 
absorbers ($N_{HI} > 10^{16}$~cm$^{-2}$). 

\section*{Fit Dependent Statistics}

\begin{figure} 
\centerline{\epsfig{file=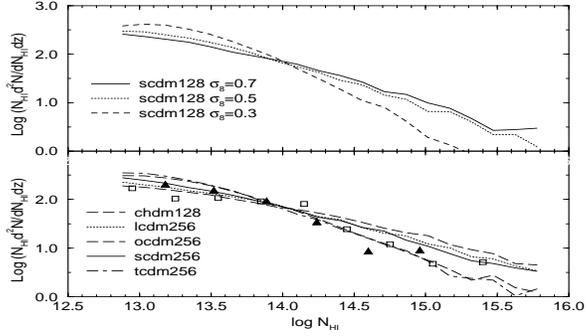,width=3.2in,height=1.9in}}
\vspace{10pt}
\caption{
HI column density distributions at $z=3$ for lines with $N_{HI}$ in the 
range $10^{12.8}-10^{16}$~cm$^{-2}$. Results for SCDM varying 
$\sigma_{8h^{-1}}$(top); results for the five cosmological models 
(SCDM, LCDM, OCDM, TCDM, CHDM) are plotted together with observed data 
from Kim, et. al~\protect\cite{kim} (open squares, $z=2.85$) and  
Kirkman \& Tytler~\protect\cite{tytler}(filled triangles, $z=2.7$)(bottom).
}\label{column}
\end{figure}

\begin{figure} 
\centerline{\epsfig{file=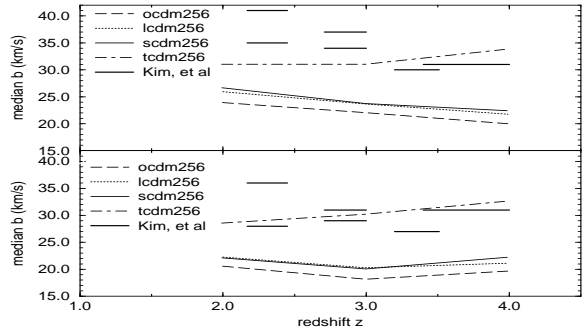,width=3.2in,height=1.9in}}
\vspace{10pt}
\caption{
The evolution of the median of the $b$ parameter distribution with redshift 
for SCDM, LCDM, OCDM, \& TCDM compared to the observed data by Kim, et. al 
\protect\cite{kim} for lines with $N_{HI}$ in the ranges 
$10^{13.8} - 10^{16}$~cm$^{-2}$ (top) and $10^{13.1}-10^{14}$~cm$^{-2}$ 
(bottom).
}\label{medb}
\end{figure}

The synthetic spectra are fit by  
Voigt profiles at these low column densities to obtain column densities and 
Doppler widths for each line.  The slope of the column density distribution 
is insensitive to changes in the size of the simulation volume or grid 
resolution. \cite{bryan}  Fig.~\ref{column} confirms analytic work 
\cite{gnedin} that the slope of the column density distribution 
depends primarily on the power in the model at scales $\sim 100 - 200$~kpc 
and steepens for models with lower power.  Each of the five cosmologies in 
our comparison study agrees with the data at the $3\sigma$ level, although 
models (SCDM,LCDM,OCDM) with moderate power at these scales are favored.
The Doppler $b$ parameter distributions are determined not only by thermal 
broadening and peculiar velocities of the absorbers, but also by the Hubble 
expansion across their width, and require high spatial 
resolution to be modeled properly. \cite{bryan,theuns} 
The shape of the  distribution including the 
high $b$ tail is well fit by hierarchical models. However, as 
shown in Fig.~\ref{medb}, observations by Kim, et. al \cite{kim} yield 
median $b$ parameters significantly higher than predicted for the models 
favored by the column density distributions (LCDM, OCDM, SCDM).

\section*{Conclusion}

\begin{figure} 
\centerline{\epsfig{file=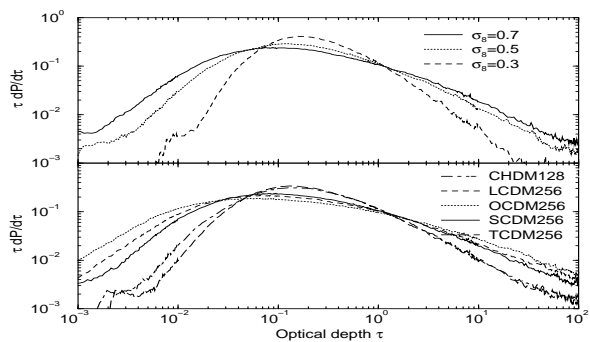,width=3.2in,height=1.9in}}
\vspace{10pt}
\caption{
Optical depth probability distribution at $z = 3$. Results for SCDM with 
varying power ($\sigma_{8h^{-1}}$) (top); results for SCDM, LCDM, OCDM, 
TCDM, \& CHDM (bottom).
}\label{tau}
\end{figure}

Although hierarchical cosmologies reproduce the general characteristics of 
the Ly$\alpha$ absorption spectra quite well, detailed tests of the models 
may require new methods of analysis for the simulations and observations.  
Statistics derived directly from the flux or optical depth distributions 
without recourse to any line fitting algorithm are particularly 
interesting.\cite{rauch} For example, the optical depth probability 
distribution, like the column density distribution, is stable to changes 
in simulation spatial resolution.  Fig.~\ref{tau} shows that the 
distribution narrows for models with lower power and that the shape of the 
distribution varies significantly for the different models over the range 
$0.05 < \tau < 4$ accessible to observations. Comparison of high quality 
observations with high resolution simulations using an ensemble of such 
statistics may soon clarify the physical properties of the intergalactic 
medium  at intermediate redshifts when galaxies were young.

This work is done under the auspices of the Grand Challenge Cosmology 
Consortium and supported in part by NSF grant ASC-9318185 and NASA 
Astrophysics Theory Program grant NAG5-3923.


\begin{references}

\bibitem{sims}
Cen, R., Miralda-Escude, J., Ostriker, J. P., \& Rauch, M. 1994, ApJ, 437, L9;
Zhang, Y., Anninos, P., Norman, M. L., \& Meiksin, A. 1997, ApJ, 485, 496;
Hernquist, L., Katz, N., Weinberg, D., \& Miralda-Escude, J. 1996, ApJ, 457,
L51

\bibitem{bryan}
Bryan, G. L., Machacek, M., Anninos, P. \& Norman, M. L. 1998, ApJ, in press 
(astro-ph/9805340)

\bibitem{machacek}
Machacek, M., Bryan, G. L., Meiksin, A., Anninos, P., Thayer, D., Norman, 
M. L., \& Zhang, Y. 1998 (in preparation)

\bibitem{bryan95}
Bryan, G. L., Norman, M. L., Stone, J. M., Cen, R., \& Ostriker, J. P. 1995, 
Comput. Phys. Comm.,89, 149 

\bibitem{anninos}
Anninos, P., Zhang, Y., Abel, T., \& Norman, M. L. 1997, New Astronomy, 2, 209

\bibitem{haardt}
Haardt, F. \& Madau, P. 1996, ApJ, 461, 20

\bibitem{gnedin}
Gnedin, N. Y. 1998, MNRAS, submitted (astro-ph/9706286)

\bibitem{theuns}
Theuns, T. Leonard, A., \& Efstathiou, G. 1998 MNRAS, 297, L49

\bibitem{kim}
Kim, T.-S., Hu, E. M., Cowie, L. L., \& Songaila, A. 1997, AJ, 114, 1

\bibitem{tytler}
Kirkman, D. \& Tytler, D. 1997, ApJ, 484, 672

\bibitem{rauch}
Rauch, M., Miralda-Escude, J., Sargent, W. L. W., Barlow, T. A., Weinberg, 
D.H., Hernquist, L., Katz, N., Cen, R., \& Ostriker, J. P. 1997, ApJ, 489, 7

\end{references}
\end{document}